\documentstyle[aps,twocolumn,epsf]{revtex}

\newcommand \beq{\begin{eqnarray}}
\newcommand \eeq{\end{eqnarray}}
\renewcommand{\theequation}{\arabic{section}.\arabic{equation}}
\def\grad{\nabla}                               
\def\del{\partial}                              
\def\bfgrad{\mbox{\boldmath$\grad$}}
\def\bfalpha{\mbox{\boldmath$\alpha$}}
\def\bfxi{\mbox{\boldmath$\xi$}}

\def\simge{\mathrel{%
   \rlap{\raise 0.511ex \hbox{$>$}}{\lower 0.511ex \hbox{$\sim$}}}}
\def\simle{\mathrel{
   \rlap{\raise 0.511ex \hbox{$<$}}{\lower 0.511ex \hbox{$\sim$}}}}

\def\bfcalA{\mbox{\boldmath${\cal A}$}}
\def\bfcalE{\mbox{\boldmath${\cal E}$}}
\def\bfcalD{\mbox{\boldmath${\cal D}$}}
\def\bfcalB{\mbox{\boldmath${\cal B}$}}

\def\journal#1#2#3#4{\ {#1}{\bf #2}, {#4} ({#3})}
\def\NPB{\journal{Nucl.\ Phys.\ {\bf B}}}

\def\PLB{\journal{Phys.\ Lett.\ {\bf B}}}

\def\PRD{\journal{Phys.\ Rev.\ {\bf D}}}
\def\PRL{\journal{Phys.\ Rev.\ Lett.}}

\def\RevModPhys{\journal{Rev.\ Mod.\ Phys.}}

\begin{document}

\title{Classical theory for non-perturbative dynamics in
hot QCD\thanks{Based on
a talk given at TFT98, Regensburg, Germany, August 10-14, 1998}}
\author{Edmond Iancu}

\address{Service de Physique Th\'eorique, CEA Saclay \\
 91191 Gif-sur-Yvette, France}

\date{\today}

\maketitle

\begin{abstract}

I present an effective classical theory which describes the
non-perturbative real-time dynamics in hot gauge theories
and has the potential for numerical implementation.

\end{abstract}
\narrowtext
\section{Introduction}
\setcounter{equation}{0}

At high temperature, the na\"{\i}ve perturbation theory 
is well known to break down at soft momenta because 
of large collective effects. 
During the last decades, there has been a considerable effort
toward the understanding of such effects and their resummation
into {\it effective} perturbative expansions,
with a better infrared behaviour.
(See Refs. \cite{BIO96,MLB} for some recent reviews
and more references.) In particular, it has been recognized,
by Braaten and Pisarski \cite{BP90}, that the gauge-invariant resummation
of the collective phenomena at the soft scale $gT$ involves 
the whole series of ``hard thermal loops'' (HTL's):  these are 
one-loop corrections to the amplitudes of the soft ($k\simle gT \ll T$) 
fields due to their scattering off the hard ($k \sim T$) plasma particles.
(See also Refs. \cite{FT90} --\cite{Kelly}.)

As well known, however, infrared (IR) problems persist even
after resumming the HTL's: in general, they are associated with (quasi)static
magnetic gluons, or photons, which are not screened by the collective
phenomena at the scale $gT$. 

In Abelian theories, we encounter only mass-shell
divergences, as, for example, in the calculation of the fermion
damping rate \cite{Pisarski93}: 
the divergences occur because a charged particle
which is nearly on-shell can emit or absorb an arbitrary number of
soft, (quasi)static, magnetic photons (the latter are virtual quanta,
so they actually describe collisions in the plasma). The kinematics is
very similar to the old ``IR catastrophe'' of zero-temperature QED \cite{BN37},
and, as a matter of facts, the divergences can be indeed removed 
\cite{lifetime,debye} by an all-orders resummation of the soft photons, 
{\it \`a la} Bloch-Nordsieck \cite{BN37}.

In QCD, the IR problems are more complicated, because of
the mutual interactions of the magnetic gluons\footnote{For definitness, 
I shall mostly use a QCD-inspired terminology. But the present considerations
apply to any hot Yang-Mills theory, in particular, to the electroweak 
plasma at high temperature ($T\gg T_{c}$).}. Perturbation theory breaks
down at the scale $g^2 T$ because of IR divergences. Specifically, 
the usual connection between
the powers of the coupling constant and the number of loops is lost:
starting with some given order in $g$ (e.g., O$(g^6)$ in the case
of the pressure \cite{Linde80}), one finds diagrams with any number of loops
which contribute at the same order. In some cases, perturbation theory 
breaks down already at leading order: this is what happens for genuinely
non-perturbative phenomena like the static magnetic screening \cite{Linde80}
or the anomalous violation of the baryon number 
in the hot electroweak theory
\cite{Shapo,Yaffe} (see also below).

For {\it static} quantities like the pressure, there exists a well defined
procedure for going beyond the perturbation theory:
this is lattice QCD at finite temperature \cite{DeTar},
possibly supplemented by dimensional reduction \cite{Kajantie,Braaten}.
But the standard lattice simulations, as formulated in imaginary
time, are not well suited for computing {\it real-time} 
correlation functions, like the baryon number violation alluded to before.
It is my purpose in this paper to describe a semi-classical approximation
which correctly reproduces the non-perturbative real-time dynamics
to lowest order in $g$ and has the potential for numerical implementation 
(a concise presentation of this method has been previously 
given in Ref. \cite{baryo}).

\section{The classical approximation}
\setcounter{equation}{0}

Let me first explain why one expects the classical approximation
to be appropriate for the problem at hand. 

The basic idea is not new  (see e.g. \cite{Shapo}):
the non-perturbative phenomena are associated with long wavelength
($\lambda \simge 1/g^2T$) magnetic fields, which, because of the
Bose-Einstein enhancement, have large occupation numbers:
\beq\label{BES}
N_0(k)\equiv\,\frac{1}{{\rm e}^{\beta k}-1}\,\simeq\,\frac{T}{k}\,
\gg 1 \qquad{\rm for}\,\,\,\, k\ll T,\eeq
and should therefore exhibit a classical behaviour.
To see this in a simple way, let us put back Planck's constant $\hbar$
in Eq.~(\ref{BES}) and compute the average energy per mode 
in thermal equilibrium:
\beq\label{EQP}
\varepsilon(k)\,=\,\frac{\hbar k}{{\rm e}^{\beta \hbar k}-1}\,\simeq\,\,{T}
\qquad{\rm as}\,\,\,\, \,\,\hbar k \ll T.
\eeq
As $\hbar \to 0$, we recover the classical equipartition theorem,
as expected. But Eq.~(\ref{EQP}) also shows that the relevant inequality
is $\hbar k \ll T$, so that the classical limit ($\hbar \to 0$ at fixed
$k$ and $T$) is actually equivalent to the soft momentum limit
($k \to 0$ at fixed $\hbar$ and $T$).

This observation is useful since we know how to perform real-time
lattice  simulations for a {\it classical} thermal field theory: 
All we have to do is to solve the classical equations of motion for given 
initial conditions, and then average over the classical phase
space with the Boltzmann weight exp$(-\beta H)$. Since the initial
conditions (say $\phi({\bf x})$ and $\dot\phi({\bf x})$ for a scalar theory)
depend only on the spatial coordinate ${\bf x}$, the phase space integration
is actually a {\it three-dimensional} functional integral, which can be
implemented on a lattice in the standard way \cite{AmbK}.

On the other hand, the classical approximation is certainly not
appropriate at high momenta, $k\simge T$, and this 
sets limits on its applicability (see below).

To be more specific, let me
consider classical Yang-Mills theory at finite temperature.
In the temporal gauge $A^0_a=0$, the independent degrees of freedom
are the vector potentials $A^i_a(x)$ and the electric fields
(``the canonically conjugate momenta'') $E^i_a(x)$,
with the Hamiltonian:
\beq\label{HYM}
H_{YM}\,=\,\frac{1}{2}\int {\rm d}^3 x\left\{{\bf E}_a\cdot{\bf E}_a\,+\,
{\bf B}_a\cdot{\bf B}_a\,\right\}.\eeq
In terms of these, the classical equations of motion read:
\beq\label{CAN0}
\del_0 A^a_i=-E^a_i,\qquad \del_0 E^a_i=\epsilon_{ijk}(D_j B_k)^a,\eeq
together with Gauss' law which in this gauge must be imposed as a constraint:
\beq\label{GAUSS0}
{\bf D\cdot E}\,=\,0.
\eeq
The thermal phase space is defined by the initial conditions
for Eqs.~(\ref{CAN0}), which I denote with calligraphic letters:
${\cal E}^a_i({\bf x})$ and ${\cal A}^a_i({\bf x})$.
Then, the canonical partition function reads:
\beq\label{ZYM}
Z_{YM}\,=\,
\int {\cal D}{\cal E}^a_i({\bf x})\,{\cal D}{\cal A}^a_i({\bf x})\,
\delta({\bfcalD\cdot \bfcalE})\,{\rm e}^{-\beta {\cal H}_{YM}},\eeq
where ${\cal H}_{YM}$ is
expressed in terms of the initial fields, cf. Eq.~(\ref{HYM}).
The classical thermal correlation functions 
can be obtained from the following generating functional:
\beq\label{Z0}
Z_{YM}[J^a_i]&=&\int {\cal D}{\cal E}^a_i{\cal D}{\cal A}^a_i\,
\delta({\bfcalD\cdot \bfcalE})\,\nonumber\\
&{}&\times\,{\exp}\left\{-\beta{\cal H}_{YM}+\int{\rm d}^4x J^a_i(x)
A^a_i(x)\right\},
\eeq
where $A^i_a(x)$ is the solution to Eqs.~(\ref{CAN0})
with the initial conditions $\{{\cal E}^a_i,{\cal A}^a_i\}$.
For instance, the free ($g=0$) two-point correlation is obtained
as:
\beq\label{D0}
D_{ij}^{0\,cl}(k)&=&(\delta_{ij}-\hat k_i\hat k_j)\,\rho_0(k_0,k)
\,N_{cl}(k_0),\nonumber\\
\rho_0(k_0,k)&\equiv& 2\pi\epsilon(k_0)\,
\delta(k_0^2 -k^2),\qquad N_{cl}(k_0)\equiv\,\frac{T}{k_0}\,,\eeq
which is simply the soft (or classical) limit 
of the corresponding Wightman functions\footnote{Recall the 
definition of the Wightman functions \cite{MLB} :
$D^>(k)\equiv\rho(k)[1+N_0(k_0)]$
and $D^<(k)\equiv\rho(k)N_0(k_0)$.}:
$D^>_0(k)\simeq D^<_0(k)\simeq (T/k_0)\rho_0(k)\equiv D_0^{cl}(k)$.
Other thermal expectation values are defined
similarly. For instance, the study of the baryon
number violation at high temperature  \cite{Shapo} requires 
the quantity $\langle (\Delta B (t))^2\rangle$, where 
\beq\label{B}
\Delta B (t)\,\propto\,
\int_0^t {\rm d}x_0\int {\rm d}^3x \,F^a_{\mu\nu} {\tilde F}_a^{\mu\nu},\eeq
(with ${\tilde F}_a^{\mu\nu} \equiv (1/2)\varepsilon^{\mu\nu\rho\lambda} 
F^a_{\rho\lambda}$) is proportional to the change in the Chern-Simons
number. And actually there have been 
attempts to compute  $\langle (\Delta B (t))^2\rangle$ via lattice
simulations of the equations above \cite{AmbK}. 

Recall, however, that the classical 
description is justified only at soft momenta (cf. Eqs.~(\ref{BES})
and (\ref{EQP})), which greatly restricts its applicability.
In general, equations like (\ref{ZYM}) or (\ref{Z0}) 
will be afflicted with ultraviolet (UV) divergences
which occur because of the replacement of the correct, quantum
distribution $N_0(k)=1/({\rm e}^{\beta k}-1)$ with
the classical one $N_{cl}(k)\equiv T/k$ (cf. Eq.~(\ref{BES})).
Recall, for instance, the famous
``ultraviolet catastrophe'' of Rayleigh and Jeans:
The classical estimate for the energy density of the black body radiation,
namely ($\Lambda$ is an ad-hoc UV cutoff):
\beq E_{cl}/V\,=\,
\int\frac{{\rm d}^3k}{(2\pi)^3}\,\varepsilon_{cl}(k)\,=\,T
\int\frac{{\rm d}^3k}{(2\pi)^3}\,\propto\,T \Lambda^3,\eeq
is obviously wrong (since UV divergent), in contrast to the quantum result:
\beq E/V\,=\,
\int\frac{{\rm d}^3k}{(2\pi)^3}\,
\frac{k}{{\rm e}^{\beta k}-1}\,=\,\frac{\pi^2 T^4}{30},\eeq
which is finite because the large momenta $k\gg T$ are exponentially 
suppressed by the Bose-Einstein thermal distribution.

Similarly, the Debye screening mass is linearly UV divergent when
computed in the classical theory. Recall,
indeed, the correct,
quantum, expression for the Debye mass to leading order in $g$
\cite{BIO96,MLB} :
\beq\label{MD}
m^2_D\,=\,-\frac{g^2 C_A}{\pi^2}\int_{0}^\infty 
{\rm d}k k^2\,\frac{{\rm d}N_0}{{\rm d}k}\,=\,\frac{g^2 C_A T^2}{3}\,,\eeq
where $C_A=N$ for SU($N$).
In the classical theory, this is replaced by:
\beq\label{MDCL}
m_{cl}^2(\Lambda) \,=\,-\,\frac{g^2 C_A}{\pi^2}\int_0^\Lambda
{\rm d}k\,k^2\,\frac{{\rm d}N_{cl}}{{\rm d}k}
\,=\,\frac{g^2 C_A T\Lambda}{\pi^2}\,,\eeq
where the UV cutoff $\Lambda$ has been introduced as an
upper momentum cutoff.

Thus, the classical approximation, as formulated above,
makes sense only for quantities which (at least, to some approximation)
are not sensitive to the hard ($k\sim T$) plasma modes.
Such quantities exist indeed:
the magnetic screening mass is an example. More generally,
all the {\it static}, non-perturbative, correlation
functions of the magnetic fields 
are correctly reproduced by the classical Yang-Mills theory, to lowest
order in $g$. (In higher orders, the quantum corrections to the
thermal distribution start to play a role.)

The easiest way to see this is to note that, in its magnetic sector, 
the classical Yang-Mills theory is the same as three-dimensional
QCD, a theory which is well known to describe the leading IR
behaviour at high-$T$ (see, e.g., \cite{Linde80,Braaten}).
Indeed, Eq.~(\ref{ZYM}) can be rewritten as:
\beq\label{ZRED0}
Z_{YM}&=&\int {\cal D}{\cal A}^a_0\,{\cal D}{\cal A}^a_i\,
\nonumber\\&{}&\,\,\times\,
\exp\left\{-\frac{\beta }{2}\int{\rm d}^3x \,\Bigl(
{\cal B}^a_i{\cal B}^a_i + ({\cal D}_i {\cal A}_0)^2
 \Bigr)\right\},\eeq
where the ${\cal A}_0^a$ components of the gauge fields have been
reintroduced as Lagrange multipliers to enforce Gauss' law.
As anticipated,  the magnetic sector in Eq.~(\ref{ZRED0})
is equivalent to three-dimensional QCD 
with dimensionful coupling constant $g^2_3 = g^2 T$.
Incidentally, this shows that the ``dimensional reduction'' 
at finite temperature (cf. \cite{Kajantie,Braaten} and Refs.
therein) is a reflection of
the classical character of the leading IR dynamics.
Thus, the various lattice investigations of QCD$_3$ in the literature
(e.g., the calculation of the magnetic mass \cite{Karsch98},
or of the non-perturbative correction, of O$(g^6)$, to the pressure
\cite{Karsch97}) can be also interpretated as classical calculations. 

On the other hand, we shall shortly see that the classical Yang-Mills
theory is not appropriate for the {\it time-dependent} non-perturbative
correlations, which remain sensitive to the hard plasma modes.
Accordingly, one expects \cite{Yaffe,Muller} the aforementioned classical
studies of the baryon number violation \cite{AmbK} not to be reliable.

\section{The role of Landau damping}
\setcounter{equation}{0}

The reason for such a failure is that the time-dependent magnetic
fields are sensitive to the hard thermal modes, via Landau
damping \cite{BIO96}. To see this, consider the propagator
of a soft magnetic gluon in the HTL approximation
(that is, after resumming the effects of the hard modes to
leading order in $g$). This reads $D_{ij}(\omega, {\bf k})
=(\delta_{ij}-\hat k_i\hat k_j)\Delta_T(\omega,k)$,
where\footnote{The retarded prescription is implicit here:
$\omega \equiv \omega + i0^+$.}:
 \beq\label{effd0}
\Delta_T(\omega,k)\,\equiv\,\frac{-1}{\omega^2-k^2 -\Pi_T(\omega,k)}\,,\eeq
and $\Pi_T(\omega,k)$ is the corresponding HTL.

In the static limit,  $\Pi_T(0,k) =0$, and there is no sensitivity
to the hard modes. More generally, all the HTL corrections vanish in
the static limit, except for the term yielding Debye screening 
\cite{BIO96,MLB}.

For non-zero $\omega$, $\Pi_T(\omega,k)$ is parametrically
of the order $m^2_D (\omega/k)$, which is of O$(g^2 T^2)$ if
$\omega \sim k$: that is, for large enough frequencies, the time-dependent
magnetic fields are screened as efficiently as the electric ones,
and decouple from the non-perturbative physics.

As $\omega \to 0$,
the polarization tensor is dominated by its imaginary part,
which describes the absorbtion of the soft gluon by a
hard thermal particle (Landau damping):
\beq\label{IMPIT}
\Pi_T(\omega \ll k)\,\simeq
\,-i\,\frac{\pi}{4}\,m_{D}^2\,\frac{\omega}{k}\,.
\eeq
Accordingly, the magnetic propagator at $\omega \ll k$ reads:
\beq\label{Vt}
\Delta_T(\omega\ll k)\simeq\,\frac{1}
{k^2-i\,(\pi \omega/4k)\,m_{D}^2},\eeq
which strongly suggests that non-perturbative phenomena can be
associated only with frequencies which are low enough: $\omega \simle
k^3/m^2_D$. For $k\sim g^2 T$, this implies $\omega \simle g^4 T$.
That is, the non-perturbative physics involves very soft
and quasistatic magnetic fields, with spatial momenta of O$(g^2 T)$
and even softer frequencies, of O$(g^4 T)$  \cite{Yaffe}.
This conclusion is further supported by the power counting analysis
in Ref.\cite{Yaffe}.

Note that the inclusion of the hard modes effects (in the form
of the HTL) was essential for the previous analysis.
In particular, the typical time scale
for non-perturbative phenomena $\tau \sim m^2_D/k^3 \sim 1/(g^4 T)$
is proportional to the Debye mass  \cite{Yaffe}, a quantity which 
is dominated by the hard modes (cf. Eq.~(\ref{MD})).
We conclude that the dynamics of the time-dependent, non-perturbative,
magnetic fields is strongly influenced by their scattering off the
hard, quantum, modes, and cannot be described within the classical
Yang-Mills theory. 

But this does not mean that the classical approximation by itself
is totally useless: we just have to be careful to apply it to the
{\it soft} fields alone. That is, we have to treat hard and soft modes
separately \cite{McLerran}: The hard ($k\sim T$) modes are quantum but
perturbative, so they can be explicitly integrated over in perturbation 
theory, to obtain an {\it effective} theory for the soft ($k\simle
gT$) modes. The resulting effective theory is non-perturbative, but for
it the classical approximation is valid, so it can be studied
numerically on a classical, three-dimensional, lattice. 

An essential step in this strategy is to clearly distinguish
between  hard and soft degrees of freedom
(to avoid overcounting, and to provide an ultraviolet cutoff
to the  effective theory for the soft modes).
Loosely speaking, this requires an intermediate scale $\Lambda$, 
with $gT \ll \Lambda\ll T$, which should act as an infrared  
cutoff for the hard modes and as an ultraviolet cutoff
for the soft ones,
and which should cancel in the calculation of physical quantities.

To lowest order in $g$, the effective theory for the soft modes
is essentially known:
this is the HTL-resummed theory by Braaten and Pisarski \cite{BP90},
which includes all the one-loop amplitudes with soft external lines
and hard loop momenta. And actually it has been already proposed by
B\"odeker, McLerran and Smilga \cite{McLerran}
to use the HTL effective theory as a classical theory for
non-perturbative calculations in real time. However,
the practical implementation of this suggestion met with several
technical difficulties: (i) There is no intermediate scale $\Lambda$ 
in the original formulation of the HTL theory: indeed, this was not 
necessary for perturbative calculations, where one can
avoid overcounting by adding and subtracting the HTL Lagrangian to,
and respectively from, the tree-level Lagrangian \cite{BP90}.
(ii) The HTL Lagrangian is non-local \cite{BIO96,MLB},
so it is a priori not obvious how to construct the associated classical 
thermal field theory (cf. the discussion in Sec. II).

Actually, problem (ii) above can be overcome by using one of
the {\it local} formulations of the HTL theory \cite{qcd,Nair}.
In what follows, I shall use the formulation in terms of kinetic equations,
as developed in Refs. \cite{qcd} (see also \cite{Kelly}). This
has the advantage to clearly emphasize the physical content
of the HTL's (see also \cite{BIO96}). Besides, we shall see later
that problem (i) above --- namely, the implementation of the
intermediate scale $\Lambda$ --- can also be solved in the framework
of the kinetic theory \cite{baryo}.

\section{Kinetic equations for HTL's}
\setcounter{equation}{0}

I consider a purely Yang-Mills plasma 
in thermal equilibrium at a temperature $T$ which is high enough 
for the coupling constant to be small: $g\ll 1$.
The typical excitations are ``hard'' gluons, 
with momenta $k\sim T$. Such gluons can develop a collective
behaviour over a typical space-time scale $\sim 1/gT$, which is large
as compared to the mean interparticle distance $\sim 1/T$.
This results in long-wavelength ($\lambda \sim 1/gT \gg 1/T$) 
colour oscillations which are most economically described
in terms of {kinetic equations} \cite{BIO96,qcd}.
In these equations, the {\it hard} ($k \sim T$) gluons are represented by
their average colour density $\delta N_a({{\bf k}},x)$
to which couple the {\it soft} (i.e., long-wavelength)
colour fields $A_a^\mu(x)$. The relevant equations read:
\beq\label{N}
(D_\nu F^{\nu\mu})_a(x)&=&
2gC_A\int\frac{{\rm d}^3k}{(2\pi)^3}\,\,v^\mu
\,\delta N_a({\bf k},x),\nonumber\\
(v\cdot D_x)_{ab}\delta N^b({{\bf k}},x)&=&-\, g\,
{\bf v}\cdot{\bf E}_a(x)\,\frac{{\rm d}N_0}{{\rm d}k}\,,\eeq
where $D^\mu=\del^\mu+igA^\mu_a T_a$ is the covariant derivative,
$E_a^i\equiv F_a^{i0}$ is the chromoelectric field,
 and $v^\mu\equiv (1,\,{\bf v})$, with ${\bf v}=
{\bf k}/k$ denoting the velocity of the hard particles ($k= |{\bf k}|$,
so that $|{\bf v}| =1$). 

The Yang-Mills equation above involves the  colour current 
induced by the hard particles:
\beq\label{j}
j^{\mu}_{a}(x)
\equiv 2gC_A\int\frac{{\rm d}^3k}{(2\pi)^3}\,v^\mu
\,\delta N_a({\bf k},x).\eeq
By solving the second Eq.~(\ref{N})
(which may be seen as a non-Abelian generalization of the familiar
Vlasov equation \cite{BIO96,Kelly,Heinz}),
we can express this current in terms of the gauge fields $A_a^\mu$,
and thus obtain an {\it effective} Yang-Mills equation which involves 
the soft fields alone:
\beq\label{ava}
D_\nu F^{\nu\mu}\,=\,m_D^2\int\frac{{\rm d}\Omega}{4\pi}\,
\frac{v^\mu v^i}{v\cdot D}\,E^i.\eeq
Here, the angular integral $\int {\rm d}\Omega$ runs over the 
orientations of ${\bf v}$, and $m_D$ is the
Debye mass, Eq.~(\ref{MD}).

Eq.~(\ref{ava}) describes the propagation of soft colour fields
in the high-$T$ plasma. The hard particles are not explicit anymore, 
since they have been integrated to yield the induced current in the r.h.s.
By expanding this current in powers of the gauge fields 
one generates \cite{qcd} all the HTL's of Braaten and Pisarski.
However, because of the  non-local structure of this current
(note the covariant derivative in the denominator), Eq.~(\ref{ava}) is
not very convenient for the construction
of the thermal partition function \cite{McLerran}.

To this aim, it is preferable to go one step back
and replace Eq.~(\ref{ava}) with the
coupled system (\ref{N}) of {\it local} equations.
There is a price to be payed for that:
in addition to the gauge fields $A^\mu_a(x)$, the local description
in Eq.~(\ref{N}) also involves the average
colour density $\delta N_a({{\bf k}},x)$,
which can be seen as an ``auxiliary field''.
Still, when working with a local theory, we are in a better position
to look for a Hamiltonian formulation, as I discuss now.

\section{The classical  effective theory}
\setcounter{equation}{0}

The first step is to recognize, on the second Eq.~(\ref{N}),
that the ${\bf v}$ and $k$-dependence 
can be factorized in $\delta N^a({{\bf k}},x)$ by writing:
\beq\label{dn}
\delta N^a({\bf k}, x)\equiv - gW^a(x,{\bf v})\,({\rm d}N_0/{\rm d}k).\eeq
The new functions $W^a(x,{\bf v})$ satisfy the equation:
\beq\label{W}
(v\cdot D_x)_{ab}W^b(x,{\bf v})\,=\,{\bf v}\cdot{\bf E}_a(x),\eeq
which is independent of $k$ since the hard particles move at the speed
of light: $|{\bf v}| =1$. 
By using Eq.~(\ref{dn}), the induced current can be written as:
\beq\label{j1}
j^\mu_a(x)&=&m_D^2\int\frac{{\rm d}\Omega}{4\pi}
\,v^\mu\,W_a(x,{\bf v}),\eeq
where the radial integration has been explicitly worked out to yield
the Debye mass (cf. Eq.~(\ref{MD})).

The Hamiltonian formulation of the effective theory \cite{baryo} involves
the auxiliary fields $W_a(x,{\bf v})$ together with the soft gauge fields
$A^\mu_a(x)$. [Note that a different Hamiltonian formulation of the 
HTL theory has been previously given by Nair \cite{Nair}, in terms of 
some other auxiliary fields.] In the temporal gauge $A^a_0=0$,
the independent degrees of freedom are 
$E^a_i$, $A^a_i$ and  $W^a$, and the corresponding equations of motion 
can be read from Eqs.~(\ref{N}), (\ref{W}) and (\ref{j1}) above.
Specifically:
\beq\label{CAN}
E^a_i&=&-\del_0 A^a_i,\nonumber\\
-\del_0 E^a_i +\epsilon_{ijk}(D_j B_k)^a &=&
m_D^2\int\frac{{\rm d}\Omega}{4\pi}\,v_i\,W^a(x,{\bf v}),\nonumber\\
\left(\del_0 + {\bf v\cdot D}\right)^{ab} W_b&=&{\bf v \cdot E}^a,\eeq
together with the constraint expressing Gauss' law:
\beq\label{GAUSS}
G^a({\bf x})\equiv
({\bf D\cdot E})^a\,+\,m_D^2\int\frac{{\rm d}\Omega}{4\pi}\,W^a(x,{\bf v})
\,=\,0.
\eeq
Note that Eqs.~(\ref{CAN}) are not in canonical form: this is already
obvious from the fact that we have an odd number of equations.
Still, it can be verified that these equations are conservative;
the corresponding, conserved energy functional has the 
gauge-invariant expression \cite{BIO96}:
\beq\label{H}
H&=&\frac{1}{2}\int {\rm d}^3 x\biggl\{{\bf E}_a\cdot{\bf E}_a\,+\,
{\bf B}_a\cdot{\bf B}_a\,\nonumber\\
&{}&\qquad\,\,+\,m_D^2
\int\frac{{\rm d}\Omega}{4\pi}\,W_a(x, {\bf v})\,W_a(x, {\bf v})\biggr\}.\eeq
Moreover, in the gauge $A^a_0=0$, the functional 
(\ref{H}) acts also 
as a Hamiltonian, that is, as a generator of the time evolution.
To see this, we need the following generalized Poisson brackets,
as first introduced by Nair\cite{Nair}:
\beq\label{PB}
\left\{E^a_i({\bf x}), A^b_j({\bf y})\right\}&=&-\,\delta^{ab}\delta_{ij}
\delta^{(3)}({\bf x-y})\,,\nonumber\\
\left\{E^a_i({\bf x}), W^b({\bf y,v})\right\}&=&v_i\,\delta^{ab}
\delta^{(3)}({\bf x-y})\,,\nonumber\\
m^2_D\left\{W^a({\bf x,v}), W^b({\bf y,v'})\right\}&=&
\Bigl(gf^{abc}W^c\,+ \nonumber\\
+\,({\bf v\cdot D}_x)^{ab} &\Bigr )&\delta^{(3)}({\bf x-y})
\delta({\bf v},{\bf v}^\prime)\,.\eeq
Here, $\delta({\bf v},{\bf v}^\prime)$ 
is the delta function on the unit sphere:
\beq
\int\frac{{\rm d}\Omega}{4\pi}\,\delta({\bf v},{\bf v}^\prime)
\,f({\bf v})\,=\,f({\bf v}^\prime),\eeq
and all the other Poisson brackets are assumed to vanish.
We also assume standard properties for such brackets, namely
antisymmetry, bilinearity and Leibniz identity. It is then
easy to verify that (a) the Poisson
brackets (\ref{PB}) satisfy the Jacobi identity (as necessary for
consistency) and (b) the equations of motion (\ref{CAN}) follow as
canonical equations for the Hamiltonian (\ref{H}). For instance,
$\del_0 W^a=\{H,W^a\}$, and similarly for $E_i^a$ and $A_i^a$.

Note also that the Hamiltonian in Eq.~(\ref{H}) is remarkably simple: 
the piece involving the auxiliary fields $W^a_0$ is simply quadratic ! 
In fact, all the dynamical complications are hidden in the
non-trivial Poisson brackets (\ref{PB}).

\section{The thermal partition function}
\setcounter{equation}{0}

We are now in position
to write down the classical partition function and compute
(generally time-dependent) thermal expectation values.
The classical phase-space is defined by the initial
conditions to Eqs.~(\ref{CAN}), and the thermal weight
is provided by the Hamiltonian in Eq.~(\ref{H}).
The correlation functions of the magnetic fields $A^i_a$ are then
generated by (compare to Eq.~(\ref{Z0})): 
\beq\label{Z}
Z_{cl}[J^a_i]&=&
\int {\cal D}{\cal E}^a_i\,{\cal D}{\cal A}^a_i\,{\cal D}{\cal W}^a\,
\delta({\cal G}^a)\,\nonumber\\&{}&\,\,\,\,\times\,
\exp\left\{-\beta {\cal H}\,+\,\int{\rm d}^4x \,J^a_i(x) A^a_i(x)\right\},\eeq
where $A^i_a(x)$ is the solution to Eqs.~(\ref{CAN})
with the initial conditions $\{{\cal E}^a_i,{\cal A}^a_i,{\cal W}^a\}$
(that is, $E^a_i(t_0,{\bf x})={\cal E}^a_i({\bf x})$, 
etc., with arbitrary $t_0$), and  ${\cal G}^a$ and ${\cal H}$
are expressed in terms of the initial fields
(cf. Eqs.~(\ref{GAUSS}) and (\ref{H})).

It can be verified \cite{baryo} that the phase-space measure 
${\cal D}{\cal E}^a_i{\cal D}{\cal A}^a_i{\cal D}{\cal W}^a$ in Eq.~(\ref{Z})
is invariant under the time evolution described by Eqs.~(\ref{CAN}),
so that $Z_{cl}[J]$ is independent of the (arbitrary)
initial time $t_0$, as it should. (This point is not trivial because of the
non-canonical structure of the equations of motion.)

Furthermore, strictly speaking, Eq.~(\ref{Z}) is not well-defined
without an UV cutoff (recall the discussion in Secs. II and III).
This will be discussed in the next two sections.
But before doing that, it might be illuminating to consider
Eq.~(\ref{Z}) in some particular cases. 

For instance, for $J=0$, it
yields the same thermodynamic potential as the leading-order dimensional
reduction \cite{Kajantie,Braaten}. Specifically
(compare to Eq.~(\ref{ZRED0})):
\beq\label{ZRED}
Z_{cl}&=&\int {\cal D}{\cal A}^a_0\,{\cal D}{\cal A}^a_i\nonumber\\
&{}&\times\exp\left\{-\frac{\beta }{2}\int{\rm d}^3x \Bigl(
{\bfcalB}^2 + ({\cal D}_i {\cal A}_0)^2
+ m_D^2{\cal A}_0^2 \Bigr)\right\},\eeq
where the fields ${\cal A}_0^a$ have been introduced
as Lagrange multipliers,  and the (Gaussian) functional
integrals over ${\cal E}^a_i$ and ${\cal W}^a$ have been explicitly 
performed. In particular, the integral over the auxiliary fields
${\cal W}^a$ has correctly generated the Debye screening mass
for the electric fields.

Consider also the Abelian limit, where the classical theory
can be exactly solved.
Physically, this describes the propagation of soft electromagnetic
waves in a hot QED plasma, with the fermions integrated over
(in the HTL approximation). In the present
approximation, the Abelian theory is quadratic, and does not require 
any UV cutoff. This is consistent with the fact that, in QED,
soft and hard degrees of freedom correspond to different kind of fields
(the hard fields are fermions, while the soft ones are photons),
so, unlike QCD, there is no danger of overcounting.

A straightforward, if lengthy, calculation  yields (see the Appendix
for details):
\beq\label{ZAB}
Z_{cl}[J_i]=\exp\left\{-\frac{1}{2}\int{\rm d}^4x{\rm d}^4y
 J_i(x) D_{ij}^{cl}(x-y) J_j(y)\right\}\eeq
with (compare to Eq.~(\ref{D0})):
\beq\label{DCL}
D_{ij}^{cl}(x-y)
\equiv\int\frac{{\rm d}^4k}{(2\pi)^4}\,{\rm e}^{-i k\cdot(x-y)}\,
\frac{T}{k_0}\,\rho_{ij}(k).\eeq
Here, $\rho_{ij}(k)$ is the photon spectral density
in the HTL approximation and in the temporal gauge:
\beq\label{RHOAX}
\rho_{ij}(k_0,{\bf k})=
(\delta_{ij}-\hat k_i\hat k_j)\rho_T(k_0,k)
\,+\,\frac{k_i k_j}{k_0^2}\,\rho_L(k_0,k),\eeq
with the (transverse and longitudinal) spectral functions
$\rho_{T,L}$ defined as standard \cite{MLB}.
They include both delta functions for on-shell quasiparticles
and off-shell pieces corresponding to Landau damping.
For instance (cf. Eq.~(\ref{effd0})):
\beq\label{RHOT}
\rho_T(k_0,k)&\equiv& 2\,{\rm Im}\,\Delta_T(k_0+i\eta,k)\nonumber\\
&=& 2\pi\epsilon(k_0) z_T(k)\delta(k_0^2 -\omega_T^2(k))
 +\beta_T (k_0,k),\eeq
where $z_T(k)$ denotes the residue of the (time-like) poles
at $k_0=\pm {\omega_T(k)}$, and $\beta_T (k_0,k) \propto
{\rm Im}\,\Pi_T(k_0,k)$ has support only at space-like momenta, 
$k_0^2 < k^2$.

More generally, in QCD, $D^{cl}$  will enter as a
propagator in the perturbative expansion of Eq.~(\ref{Z}).

\section{The hard-soft matching}
\setcounter{equation}{0}

As already mentioned,  Eq.~(\ref{Z}) must be supplemented
with an UV cutoff $\Lambda$ which is meant to separate
between hard and soft gluons: $gT \ll \Lambda \ll T$.
Correspondingly, the effective Hamiltonian must include some 
$\Lambda$-dependent counterterms, chosen so as to cancel the dependence
on $\Lambda$ in the calculation of physical quantities
(a cancellation to be referred to as ``matching''). 
The matching can be done only to one-loop order (since
the HTL theory is already an one-loop approximation). But this 
is enough to remove the most dangerous, linear, UV divergences of the 
classical theory (see below).

For more clarity, consider first a very simple cutoff procedure:
the UV cutoff $\Lambda$ is implemented as an upper momentum 
cutoff on the loop integrals of the classical theory\footnote{By 
``loops'', I mean here, of course, the {\it classical},
thermal, loops, as generated in Eq.~(\ref{Z}) by
the non-linearities of the classical theory.}.
(This would be just fine if we were interested in 
 perturbative calculations.)
For matching purposes, we need the $\Lambda$-dependent 
corrections to the soft amplitudes, to one-loop order
in the effective theory. Such corrections are easy to isolate
\cite{baryo}: they are precisely the ``HTL's'' of the classical theory.
Namely, they differ from the usual HTL's only by the replacement
of the correct Debye mass $m_D^2$,  Eq.~(\ref{MD}),
 with the $\Lambda$-dependent classical
``Debye mass'' $m^2_{cl}(\Lambda)$, Eq.~(\ref{MDCL}).

In this case, the matching is very easy: just
replace the  physical Debye mass $m_{D}^2$ in the definition of the
effective theory (that is, in all the equations in Secs. V and VI)
with the following $\Lambda$-dependent parameter: 
\beq\label{MDMU}
m^2_H(\Lambda) \equiv m_D^2-m_{cl}^2(\Lambda)\,=\,
\frac{g^2C_A T^2}{3}
\left(1\,-\,\frac{3}{\pi^2}\frac{\Lambda}{T}\right).\eeq
Then, to one-loop order, the effective theory generates the correct
soft amplitudes, that is, the usual HTL's with total strength
$m^2_H(\Lambda)+m_{cl}^2(\Lambda)=m_{D}^2$.

For non-perturbative calculations, however, one cannot use a simple
momentum cutoff, as above, but one rather needs an UV cutoff which 
can be also implemented on the lattice. This
cannot be the lattice spacing itself: indeed, a finite
(and relatively large) lattice spacing $a\sim 1/\Lambda \gg 1/T$
will introduce lattice artifacts (essentially, the loss
of rotational symmetry) which will make impossible the matching 
with the hard sector \cite{McLerran,Arnold}.

A first solution to this problem have been proposed in
Ref. \cite{baryo}: there, a smooth
cutoff has been introduced in the classical theory
by replacing, in Eq.~(\ref{H}),
\beq\label{REG}
{\rm Tr}\, B^i B^i\,\longrightarrow\,{\rm Tr}\, B^i f\left(\frac
{{\bf D}^2}{\Lambda^2}\right) B^i,\eeq
where $f(z)=1+z^2$ and the trace refers to color indices.
In principle, such a higher-derivative cutoff can be indeed
implemented on the lattice, as discussed in Ref. \cite{Arnold}. 
Then, the matching can be done (at least for the non-perturbative 
correlation functions) by adding a $\Lambda$-dependent counterterm to the
Debye mass, as in Eq.~(\ref{MDMU}) \cite{baryo}.
In practice, however, this might be quite tedious,
as it requires highly improved lattice 
Hamiltonians \cite{Arnold}. Fortunately, there exists
a simpler way to perform classical non-perturbative
calculations without running into ultraviolet problems.
This will be presented in the next section.

\section{The easy way}
\setcounter{equation}{0}

Note first that, within the effective theory, it is easy to
distinguish between the gluon fields which are responsible
for the non-perturbative IR phenomena and those which are
responsible for the leading UV behaviour:

The non-perturbative phenomena are generated by soft
and quasistatic magnetic gluons  (cf. Sec. III),
which are described by the {\it off-shell} piece $\propto \beta_T$
of the gluon propagator (cf. Eqs.~(\ref{DCL}) and (\ref{RHOT})).

The UV behaviour, on the other hand, is determined by the
relatively hard ($k\sim \Lambda$) magnetic gluons,
which are essentially {\it on-shell} (and, actually, even on
the {\it tree-level} mass shell). Indeed, for momenta 
$k \sim \Lambda \gg m_D$, the HTL corrections are negligible
and the classical propagator in Eq.~(\ref{DCL}) reduces to its 
tree-level expression, Eq.~(\ref{D0}).
In fact, the ``classical HTL's'' mentioned in Sec. VII 
(in relation with the matching problem) have been computed 
\cite{McLerran,baryo,Arnold} by using the free propagator
$D_0^{cl}(k)$ for the hard ($k\sim \Lambda$) gluons around the loop.

This suggests that one can avoid the UV problems
of the classical theory by subtracting
(in a way to be specified shortly) the tree-level piece $\rho_0(k)$,
Eq.~(\ref{D0}), from the magnetic spectral density
$\rho_T(k)$, Eq.~(\ref{RHOT}).

To understand how this can be achieved in practice, 
recall from Sec. VI that the magnetic two-point function
$D_T^{cl}(k)\equiv(T/k_0)\rho_T(k)$ is obtained by averaging
the classical solution over the initial conditions
$\{{\cal A}^i,{\cal E}^i, {\cal W}\}$ at the (arbitrary)
initial time $t_0$. By using the equations in the Appendix,
it can be verified that, as $t_0\to -\infty$, 
(i) the {\it off-shell}
piece of the propagator, proportional to
$\beta_T$, is generated by the
integral over the colour fluctuations ${\cal W}_a$, while
(ii) the {\it on-shell} piece arises from the integration over the
initial colour fields ${\cal A}^i_a$ and ${\cal E}^i_a$.

This is physically intuitive: (i) the off-shell piece describes Landau
damping, so it should arise from the coupling of the soft fields
to the long-range colour fluctuations ${\cal W}_a$ of the hard particles;
(ii) at high momenta $k\sim\Lambda \gg m_D$,
the one-shell piece goes into the tree-level propagator $D_0^{cl}(k)$,
which is the same as in the purely Yang-Mills theory
(cf. Eqs.~(\ref{Z0}) and (\ref{D0})), and is therefore independent
of the ${\cal W}$'s.

This can be also verified by a rapid calculation
(see the Appendix for more details):
To this aim, it is enough to consider the
linearized (or Abelian) version of the effective theory,
since this is what determines the propagator.
The corresponding equation for the transverse
field $A_T^i(x)$ reads, for $x_0> t_0$ (cf. Eq.~(\ref{MAXT})):
\beq\label{MAXT0}
(\del_0^2 - \bfgrad^2)A_T^i+
\int_{t_0}^\infty {\rm d}^4y\,
\Pi_T(x-y)A^i_T(y)=\xi_T^i(x),\eeq
where $\xi^i$ is determined by the
initial fluctuations ${\cal W}({\bf x,v})$,
cf. Eq.~(\ref{XIA}):
\beq\label{XI0}
\xi^i(x)\,=\,m_D^2\int\frac{{\rm d}\Omega}{4\pi}\,v^i
\,{\cal W}({\bf x-v}(x_0-t_0),{\bf v}),\eeq
If $t_0\to -\infty$, Eq.~(\ref{MAXT0}) can be easily 
solved by Fourier transform. For vanishing initial conditions,
${\cal A}_T^i={\cal E}_T^i=0$, the solution reads:
\beq\label{ATT}
A_T^i(k)=\Delta_T(k)\,\xi_T^i(k),\eeq
where $\Delta_T(k)$ is the retarded magnetic propagator
in Eq.~(\ref{effd0}). The correlation function:
\beq
\langle A_T^i(k) A_T^{j\,*}(p)\rangle 
\,\equiv\, (2\pi)^4 \delta^{(4)}(p+k)\,\tilde D_T^{cl}(k) \eeq
can then be obtained by performing the functional average
over the ${\cal W}$'s. Equivalently, note that, according to
Eqs.~(\ref{Z}) and (\ref{H}), the ${\cal W}$'s can be seen as Gaussian 
random variables with the local correlation function:
\beq\label{WN0}
\langle {\cal W}({\bf x, v})\,{\cal W}({\bf y, v}^\prime)\rangle=
(T/m^2_D)\,\delta^{(3)}({\bf x-y})\delta({\bf v},{\bf v}^\prime).
\eeq
This immediately implies:
\beq\label{XIXI}
\langle \xi^i(x) \xi^j(y)\rangle&=&m^2_D T
\int\frac{{\rm d}\Omega}{4\pi} v^i v^j
\delta^{(3)}({\bf x-y}-{\bf v}(x_0-y_0)),\nonumber\\
&{}&\eeq
and therefore, after some simple algebra,
\beq\label{DT}
\tilde D_T^{cl}(\omega,k)&=&- 2\,(T/\omega)\,|\Delta_T(k)|^2\,\,
{\rm Im} \,\Pi_T(\omega,k)\nonumber\\
&\equiv& (T/\omega)\, \beta_T(\omega,k),\eeq
where in writing the last equation I have recognized the 
imaginary part of the HTL polarization tensor \cite{BIO96}:
\beq
{\rm Im}\,\Pi^{ij}(\omega, k)\,=\,-\pi \omega 
m_D^2\int\frac{{\rm d}\Omega}{4\pi}\,
v^i v^j \delta(\omega-{\bf v \cdot k})\,,\eeq
together with the definition (\ref{RHOT}) of the off-shell spectral 
function $\beta_T$. Eq.~(\ref{DT}) is precisely the Landau damping piece
of the magnetic propagator in Eqs.~(\ref{DCL})--(\ref{RHOT}).
As anticipated, this has been generated here by averaging over
the initial fluctuations ${\cal W}$, while keeping
constant the initial fields ${\cal A}_T^i$ and ${\cal E}_T^i$.

The above discussion suggests the following strategy for
non-perturbative calculations in
the classical effective theory:
The (fully non-linear) equations of motion (\ref{CAN})
should be solved with the following initial conditions at $t=t_0$:
\beq\label{INIT0}
{\bfcalA}^a({\bf x})\,=\,0,\qquad {\bfcalE}_T^a({\bf x})\,=\,0,\qquad
{\cal W}^a({\bf x,v}),\nonumber\\
ik {\cal E}_L^a(k) \,=\,m_D^2\int\frac{{\rm d}\Omega}{4\pi}\,
{\cal W}^a({\bf k, v})\,.\eeq
That is, the initial colour fluctuations ${\cal W}^a$ are arbitrary,
but all the initial gauge fields are taken to be zero, except for the
longitudinal component of the electric field which is fixed 
by Gauss' law (\ref{GAUSS}).
The system is let to evolve for a long time (so as to simulate
the limit $t_0\to -\infty$), and then the correlation function of interest
(e.g., Eq.~(\ref{B})) is evaluated along the classical solution.
This yields the correlation function as a functional of the ${\cal W}$'s,
say $\Omega[{\cal W}]$, which is then averaged as follows
(below, ${\cal N}\equiv 1/\langle 1 \rangle$): 
\beq\label{AVOM}
\langle \Omega \rangle &\equiv& {\cal N}
\int {\cal D}{\cal E}_L{\cal D}{\cal W}\,\delta({\cal G})\,
\Omega[{\cal W}]\,\nonumber\\&{}&\,\,\times\,
\exp{\Bigl\{-\frac{\beta}{2}\int {\rm d}^3 x\Bigl(
{\cal E}_L^2 +m_D^2
\int\frac{{\rm d}\Omega}{4\pi}\,{\cal W}^2\Bigr)\Bigr\}}.\eeq
Clearly, this equation involves only a {\it restricted} phase-space
integration (compare to Eq.~(\ref{Z})), in the sense that the
initial transverse fields are kept constant (and equal to zero).
But this is precisely what we need in order to generate the correct IR 
physics, without being sensitive to the hard classical modes. 

More precisely, the perturbative expansion
of Eq.~(\ref{AVOM}) will involve the {\it reduced} transverse
propagator $\tilde D_T^{cl}$, Eq.~(\ref{DT}), 
rather than the full classical propagator in Eq.~(\ref{DCL}).
The one-loop classical amplitudes constructed with
$\tilde D_T^{cl}$ will be UV finite, since dominated by soft
loop momenta $k\simle gT$. For instance, the tadpole diagram 
constructed from  $\tilde D_T^{cl}$ reads:
\beq \tilde\Pi_{tp}&\propto&
\int\frac{{\rm d}^4k}{(2\pi)^4}\,\tilde D_T^{cl}(k)\,=\,
T \int\frac{{\rm d}^3k}{(2\pi)^3}
\int\frac{{\rm d}\omega}{2\pi \omega}\,\beta_T(\omega,k)
\nonumber\\&=& T \int\frac{{\rm d}^3k}{(2\pi)^3}
\left(\frac{1}{k^2}\,-\,\frac{z_T(k)}{\omega_T^2(k)}\right),\eeq
which is indeed finite (and of O$(m_D T)$)
since the integrand in the last equation
behaves like $m_D^2/k^4$ at momenta $k\gg gT$ \cite{lifetime}.
(To obtain this last equation, I have used the sum rule (\ref{SR})
for $\rho_T/\omega$, together with Eq.~(\ref{RHOT}).)
This should be contrasted with the usual (classical)
tadpole, as generated by the full propagator $D_T^{cl}$,
which is linearly UV divergent:
\beq \Pi_{tp}&\propto&
\int\frac{{\rm d}^4k}{(2\pi)^4}\, D_T^{cl}(k)\,=\,
T \int\frac{{\rm d}^3k}{(2\pi)^3}
\int\frac{{\rm d}\omega}{2\pi \omega}\,\rho_T(\omega,k)\, 
\nonumber\\&=& T \int\frac{{\rm d}^3k}{(2\pi)^3}\,\frac{1}{k^2}\,
\sim \Lambda T\,.\eeq
In general, all the ``classical HTL's'' of Sec. VII will be identically
zero when computed with the reduced propagator $\tilde D_T^{cl}$,
so there is no need for matching.

\begin{figure}
\protect \epsfysize=7.cm{\centerline{\epsfbox{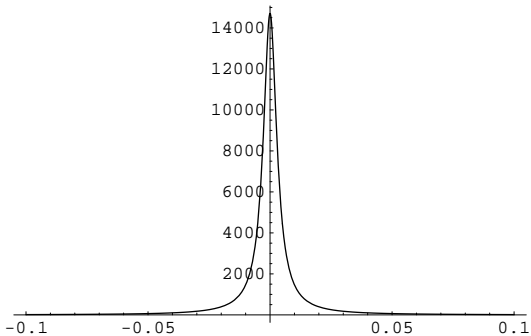}}}
	\caption{The quantity $\rho_T(\omega,k)/\omega$ as a function
of $\omega$ for $k=0.1\,m_D$.  All the quantities are made adimensional
by multiplying them by appropriate powers of $m_D$.}
\label{rho2}
\end{figure}

On the other hand, the reduced propagator generates the correct
infrared structure, as determined by the
off-shell, soft ($k\sim g^2T$) and quasistatic
($\omega\simle g^4T$) magnetic gluons, 
In fact, the spectral weight of the non-perturbative
gluons is concentrated at low frequencies $\omega\ll k$,
so it is entirely contained in $\tilde D_T^{cl}$. 
To see this, recall the low frequency behaviour
of $\beta_T/\omega$ (cf. Eq.~(\ref{Vt})):
\beq\label{rhot1}
\frac{\beta_T(\omega\ll k)}{\omega}\,\,\simeq\,\, \frac{\pi}{2}\,
 \frac{m_{D}^2 \, k}
{k^6\,+\, (\pi m_{D}^2 \omega/4)^2}\,.\eeq
For $k\ll m_D$, this is strongly peaked at $\omega=0$,
with a width $\Delta \omega \simeq k^3/m^2_D$.
(See Fig.~\ref{rho2} for an illustration.) For $k\sim g^2 T$,
this yields $\omega\simle g^4T$. In particular, as $k \to 0$,
$\beta_T(\omega, k)/\omega\to (2\pi/k^2)\delta(\omega)$,
which is just another way to see the ``dimensional reduction''
\cite{lifetime}.

\section{Conclusions}
\setcounter{equation}{0}

I have presented here a classical field theory for hot QCD which
describes the non-perturbative infrared physics to lowest
order in $g$ and is well suited for numerical simulations on a lattice.
This is an effective theory for the soft gluons (with momenta
$k\simle gT$), as obtained after integrating out the hard ones
($k\sim T$) in the HTL approximation (i.e., to leading order in $g$).
The classical theory has a built-in UV cutoff $\Lambda$
(with $gT\ll \Lambda \ll T$) which can be introduced as a 
higher-derivative cutoff for lattice purposes.
In practice, there is a simpler way to avoid UV problems, which is
to restrict the phase-space averaging to the colour fluctuations
of the hard gluons (the ``auxiliary fields'' $W_a(x,{\bf v})$). 

The classical theory allows for three-dimensional lattice simulations
of the real-time dynamics. In this sense, it appears as a generalization 
of the (leading-order) dimensional reduction (DR), to which it reduces
in the static limit (cf. Eq.~(\ref{ZRED})). However --- and
unlike DR which, at least in principle, can be pushed to arbitrarily
high orders \cite{Kajantie,Braaten} ---, the present procedure is
{\it not} systematic.
That is, one cannot improve the classical theory by integrating
out the hard modes, e.g., to next-to-leading order in $g$,
not only because the latter operation would be technically involved,
but also because such an improvement would not be consistent
with the validity of the classical approximation. Different to say,
the quantum corrections to the dynamics of the soft fields will
generally enter on the same footing as the hard-field corrections
to the HTL's. Besides, the whole construction of the classical 
theory as given above has relied in a very non-trivial way on the
specific structure and symmetries of the HTL's.

Still, the effective theory formulated as above is accurate enough
to allow for leading-order calculations (e.g, on a lattice) of
non-perturbative real-time correlations like the hot baryon number
violation rate \cite{Shapo,Yaffe,AmbK,Muller}, the quasiparticle
damping rates \cite{Pisarski93,lifetime}, or some transport
coefficients (e.g., the bulk viscosity \cite{JY96}).

The effective theory should be further compared to the
method developed in Refs. \cite{Muller}, where classical coloured
particles have been introduced to simulate the HTL's
on the lattice (see also Refs. \cite{Kelly,Heinz}).
The present formalism has the advantage to avoid the UV problem
of the classical theory, thus allowing for the continuum
limit $a\to 0$ to be eventually taken.
Moreover, this formalism should also include, as a special limit,
the effective theory recently proposed by B\"odeker \cite{Bodeker}.
Therefore, by performing accurate lattice calculations in the present
formalism, it should be possible to improve over the calculation in Ref. 
\cite{Muller}, and eventually distinguish between the various theoretical
predictions \cite{Shapo,Yaffe,Bodeker} for the hot
baryon number violation rate.

\setcounter{equation}{0}
\vspace*{.7cm}
\renewcommand{\theequation}{A.\arabic{equation}}
\appendix{\noindent {\large{\bf Appendix A}}}
\vspace*{.2cm}

Below, I shall derive Eqs.~(\ref{ZAB})--(\ref{DCL}) for
the thermal partition function in QED.
The first step is to solve the Abelian version of Eqs.~(\ref{CAN}),
namely ($E^i\equiv -\dot A^i$):
\beq\label{EAB}
\Bigl[(\del_0^2 - \bfgrad^2)\delta^{ij}+\del^i\del^j\Bigr]
A^j(x)&=&m_D^2\int\frac{{\rm d}\Omega}{4\pi}\,v^i\,W(x,{\bf v}),\nonumber\\
\left(\del_0 + {\bf v}\cdot\bfgrad \right)W(x,{\bf v})&=&{\bf v \cdot E}
(x),\eeq
with the initial conditions at $t_0=0$:
\beq\label{INIT}
A^i(0,{\bf x})&=&{\cal A}^i({\bf x}),\qquad
\dot A^i(0,{\bf x})\,=\, -{\cal E}^i({\bf x}),\,\,\nonumber\\
W(0,{\bf x,v})&=&{\cal W}({\bf x,v}),\eeq
which are constrained by Gauss' law:
\beq\label{GAB}
\bfgrad \cdot \bfcalE \,- \,m_D^2\int\frac{{\rm d}\Omega}{4\pi}\,
{\cal W}({\bf x, v})\,=\,0.\eeq
To completely fix the gauge, I further restrict the initial
fields to be tranverse: $\bfgrad \cdot \bfcalA =0$.

We thus have a linear initial value problem which can be solved 
by Laplace transform, in the standard way. (See, e.g.,
Ref. \cite{Boyan} for a similar problem.)

Consider first the Vlasov equation (i.e., the second equation
(\ref{EAB})) for $W(x,{\bf v})$. Its solution is conveniently
written as:
\beq\label{DA}
W\,=\,W_{ind}+W_{fl},\eeq
where $W_{ind}$ is the piece induced by the gauge fields,
i.e., the solution to the Vlasov equation with zero
initial condition ($W_{ind}(0,{\bf x,v})=0$),
while  $W_{fl}$ satisfies the homogeneous equation:
\beq\label{eqWFL}
\left(\del_0 + {\bf v}\cdot\bfgrad \right)W_{fj}&=&0,\eeq
with the initial condition $W_{fl}(0,{\bf x,v})
={\cal W}({\bf x,v})$.
It immediately follows that:
\beq\label{WFLA}
W_{fl}(x,{\bf v})\,=\,{\cal W}({\bf x-v}x_0,{\bf v}),\eeq
which is referred to as ``the fluctuating piece'' since it is
determined by the initial charge fluctuations ${\cal W}$.
Also, for $x_0 > t_0\,$:
\beq\label{WIA}
W_{ind}(x,{\bf v})&=&\int {\rm d}^4 y\,\theta(y_0)\,G_R(x,y|{\bf v})\,
\,{\bf v \cdot E}(y),\eeq
where $G_{R}(x,y|{\bf v})$ is the retarded 
Green's function of the line derivative
$\del_0 + {\bf v}\cdot\bfgrad\,$:
\beq\label{GR0}
G_{R}(x,y|{\bf v})=\theta (x_0-y_0)\,\delta^{(3)}
\left({{\bf x}}-{{\bf y}}-{{\bf v}}(x_0-y_0)
\right ).\eeq
Eq.~(\ref{DA}) implies a similar decomposition for the current
(\ref{j1}): $j^i=j^i_{ind} + \xi^i$, where  $\xi^i$ is the
fluctuating piece:
\beq\label{XIA}
\xi^i(x)\,=\,m_D^2\int\frac{{\rm d}\Omega}{4\pi}\,v^i
\,{\cal W}({\bf x-v}x_0,{\bf v}),\eeq
and $j^i_{ind}$ is the current determined by $W_{ind}\,$:
\beq\label{jind}
j^i_{ind}(x)&\equiv&m_D^2\int\frac{{\rm d}\Omega}{4\pi}
\,v^i\,W_{ind}(x,{\bf v})\nonumber\\
&=&-\int {\rm d}^4 y\,\theta(y_0)\,\Pi^{ij}(x-y)A^j(y)\nonumber\\
&{}&+\,m_D^2\int\frac{{\rm d}\Omega}{4\pi}\,v^i v^j
{\cal A}^j({\bf x-v}x_0),\eeq
where in writing the second line I have used Eq.~(\ref{WIA})
together with an integration by parts, and I have identified the
(retarded) polarization tensor in the HTL approximation:
\beq\label{Pi}
\Pi^{ij}(x-y)&=&m_D^2 \int\frac{{\rm d}\Omega}{4\pi}\,v^i v^j\,
\del_0G_R(x,y|{\bf v})\nonumber\\&=&
\int\frac{{\rm d}^4k}{(2\pi)^4}\,{\rm e}^{-ik\cdot(x-y)}
\Pi^{ij}(k),\eeq
with the familiar HTL expression \cite{BIO96}:
\beq
\Pi^{ij}(k)\,=\, m_D^2\, k_0\int\frac{{\rm d}\Omega}{4\pi}\,
\frac{v^i v^j}{k_0-{\bf v \cdot k} +i\eta}\,.\eeq
The Maxwell equation (i.e., the first  equation
(\ref{EAB})) reads then (for $x_0 > t_0$):
\beq\label{MAX}
\Bigl[(\del_0^2 - \bfgrad^2)\delta^{ij}+\del^i\del^j\Bigr]
A^j(x)&=&j^i_{ind}(x) + \xi^i(x),\eeq
and can be most easily solved by going to the three-momentum
space and then separating the Fourier modes into
transverse and longitudinal components:
\beq
A^i(t,{\bf x})&=&
\int\frac{{\rm d}^3k}{(2\pi)^3}\,{\rm e}^{i{\bf k\cdot x}}
A^i(t,{\bf k}),\nonumber\\
A^i(t,{\bf k})&=&A_T^i(t,{\bf k}) + {\hat k}^i A_L(t,k),\eeq
where ${\bf k \cdot A}_T = 0$. A similar decomposition holds
for the initial fields ${\cal A}^i$ and  ${\cal E}^i$,
with  ${\cal A}_L =0$ (gauge fixing) and ${\cal E}_L$ as determined
by Gauss' law (cf. Eq.~(\ref{GAB})):
\beq\label{GAB1}
ik {\cal E}_L(k) \,=\,m_D^2\int\frac{{\rm d}\Omega}{4\pi}\,
{\cal W}({\bf k, v})\,.\eeq

Let me consider the transverse sector in more detail.
Eqs.~(\ref{MAX}) and (\ref{jind}) yield:
\beq\label{MAXT}
(\del_0^2 +k^2)A_T^i(t)&+&
\int {\rm d}t'\,\theta(t')\Pi_T(t-t')A^i_T(t')\,-\nonumber\\
&{}&\qquad -\,\alpha^i(t)\,=\,\xi_T^i(t),\eeq
where the ${\bf k}$-dependence is implicit, $\xi_T^i(t,{\bf k})$
is the transverse projection of $\xi^i$, Eq.~(\ref{XIA}), and 
\beq
\alpha^i(t,{\bf k})\,\equiv\,
m_D^2\int\frac{{\rm d}\Omega}{4\pi}\,{\rm e}^{-i{\bf v\cdot k}t}v^i v^j
{\cal A}_T^j({\bf k}),\eeq
which is transverse as well: ${\bf k}\cdot \bfalpha=0$
(N.B. ${\cal A}^i\equiv {\cal A}_T^i\,$.)

The solution to Eq.~(\ref{MAXT}) corresponding to
the initial conditions (\ref{INIT}) can be written as in
Eq.~(\ref{DA}):
\beq\label{ATA}
{\bf A}_T(t,{\bf k})\,=\,{\bf A}_T^{ind}(t,{\bf k})
\,+\,{\bf A}_T^{fl}(t,{\bf k}),\eeq
where ${\bf A}_T^{ind}$ is the piece determined by the
initial electromagnetic fields $\{{\cal A}_T^i,{\cal E}_T^i\}$,
while ${\bf A}_T^{fl}$ is generated
by the ${\cal W}$'s (via $\xi_T^i$). Specifically:
\beq\label{AINDA}
{\bf A}_T^{ind}(t,{\bf k})=
\int\frac{{\rm d}\omega}{2\pi}\rho_T(\omega)\Bigl\{
k^2\bfcalA_T\frac{\cos{\omega t}}{\omega}-\bfcalE_T{\sin{\omega t}}
\Bigr\}\eeq
and, respectively,
\beq\label{AFLA}
{\bf A}_T^{fl}(t,{\bf k})
&=&\int {\rm d}t'\,\theta(t')\,\Delta_T(t-t')\,\bfxi_T(t')\,,\eeq
where $\Delta_T$ is the (retarded) transverse propagator
(cf. Eq.~(\ref{effd0})), with spectral density
$\rho_T$ (cf. Eq.~(\ref{RHOT})).

Eqs.~(\ref{ATA}) to (\ref{AFLA}) express the transverse solution
${\bf A}_T(t,{\bf k})$ as an explicit functional of the initial
conditions (\ref{INIT}) (recall Eq.~(\ref{XIA})).
The {\it fluctuating} piece in Eq.~(\ref{AFLA}) is the solution to
the inhomogeneous equation (\ref{MAXT}) with zero initial conditions:
${\cal A}_T^i={\cal E}_T^i=0$. The {\it induced} piece, Eq.~(\ref{AINDA}), 
satisfies the homogeneous equation (i.e., Eq.~(\ref{MAXT})
with $\xi_T^i=0$) with
the proper initial conditions  $\{{\cal A}_T^i,{\cal E}_T^i\}$.
To verify the latter, use the following sum-rules \cite{Pisarski93,MLB} :
\beq\label{SR}
\int\frac{{\rm d}\omega}{2\pi}\,\omega\rho_T(\omega)\,=\,1\,=\,
k^2\int\frac{{\rm d}\omega}{2\pi}\,\frac{\rho_T(\omega)}{\omega}.\eeq

The next step is to perform the average over the initial conditions,
as expressed by the following functional integral (cf. Eq.~(\ref{Z})):
\beq\label{ZT}
Z_{cl}[{\bf J}_T]&=&
\int {\cal D}{\cal E}^i{\cal D}{\cal A}^i{\cal D}{\cal W}\,
\delta({\cal G})\,
{\rm e}^{-\beta {\cal H}+\int{\rm d}^4x {\bf J}_T\cdot{\bf A}_T}
\nonumber\\ &=& Z^{ind}[{\bf J}_T]\times Z^{fl}[{\bf J}_T],\eeq
where (recall that ${\cal A}_L=0$):
\beq\label{ZIF}
Z^{ind}[{\bf J}_T]&\equiv&
\int {\cal D}{\cal E}_T^i{\cal D}{\cal A}_T^i\,
{\rm e}^{-\beta {\cal H}_T+\int{\rm d}^4x {\bf J}_T\cdot{\bf A}_T^{ind}},
\nonumber\\
Z^{fl}[{\bf J}_T]&\equiv&
\int {\cal D}{\cal E}_L{\cal D}{\cal W}\,\delta({\cal G})\,
{\rm e}^{-\beta {\cal H}_L+\int{\rm d}^4x {\bf J}_T\cdot{\bf A}_T^{fl}},\eeq
and I have denoted:
\beq\label{HLT}
{\cal H}_T&\equiv&\frac{1}{2}\int {\rm d}^3 x\Bigl(
{\bfcalE}_T^2 + {\bfcalB}^2\Bigl)\nonumber\\
{\cal H}_L&\equiv&\frac{1}{2}\int {\rm d}^3 x\Bigl(
{\cal E}_L^2 +m_D^2
\int\frac{{\rm d}\Omega}{4\pi}\,{\cal W}^2\Bigr),\eeq
with ${\cal H}={\cal H}_T+{\cal H}_L$.

The first integral in Eq.~(\ref{ZIF}) is straightforward and yields (cf. 
Eq.~(\ref{AINDA})):
\beq\label{ZIND}
Z^{ind}[{\bf J}_T]&=&
\exp\biggl\{\frac{T}{2}
\int\frac{{\rm d}^3k}{(2\pi)^3}\int \frac{{\rm d}\omega}{2\pi}
\frac{{\rm d}\omega'}{2\pi}\,\,{\bf J}_T(\omega)
\cdot {\bf J}^*_T(\omega')\nonumber\\
&{}&\qquad\,\times\,\Bigl(1+\frac{k^2}{\omega\omega'}\Bigr)
\rho_T(\omega,k)\rho_T(\omega',k)\biggr\}.\eeq
The second integral involves ${\bf A}_T^{fl}$ which
is proportional to ${\cal W}$. Specifically, Eqs.~(\ref{XIA}) and
(\ref{AFLA}) imply:
 \beq\label{AFL}
{\bf A}_T^{fl}(\omega,{\bf k})&=& - 2
m_D^2\int\frac{{\rm d}\Omega}{4\pi}\,\,{\bf v}_T\,{\cal W}
({\bf k, v})\nonumber\\&{}&\qquad\qquad\times\,
{\rm Im}\,\,\frac{\Delta_T^R(\omega,k)}{\omega
-{\bf v\cdot k}+i\eta}\,.\eeq
By using this expression, together with the identity:
\beq\label{IDENT}
\int {\cal D}{\cal E}_L{\cal D}{\cal W}\,\delta({\cal G})\,
{\rm e}^{-\beta {\cal H}_L+\int{\rm d}^3x
\int\frac{{\rm d}\Omega}{4\pi}\,{\cal K}{\cal W}}\,=\qquad\nonumber\\
=\,\exp\biggl\{\frac{T}{2 m^2_D}
\int\frac{{\rm d}^3k}{(2\pi)^3}\int \frac{{\rm d}\Omega}{4\pi}
\int \frac{{\rm d}\Omega'}{4\pi}\,{\cal K}({\bf k, v})\nonumber\\
\,\,\times\,\Bigl[\delta({\bf v,v'})-\frac{m_D^2}{m_D^2 +k^2}\Bigl]
{\cal K}^*({\bf k, v'})\biggr\},\eeq
one eventually obtains:
\beq\label{ZTA}
Z_{cl}[{\bf J}_T]&\equiv& Z^{ind}[{\bf J}_T]\times Z^{fl}[{\bf J}_T]
\nonumber\\&=&
\exp\biggl\{\frac{T}{2}
\int\frac{{\rm d}^4k}{(2\pi)^4}\,\frac{\rho_T(k_0,k)}{k_0}\,
\,|{\bf J}_T(k)|^2
\biggr\}.\eeq
As expected, this provides the {\it transverse} piece of the
classical two-point function in Eqs.~(\ref{DCL})--(\ref{RHOAX}).

The corresponding {\it longitudinal} piece follows from:
\beq\label{ZL}
Z_{cl}[J_L]&=&
\int {\cal D}{\cal E}_L{\cal D}{\cal W}\,\delta({\cal G})\,
{\rm e}^{-\beta {\cal H}_L+\int{\rm d}^4x\, J_L A_L},\eeq
where $A_L\equiv A_L^{fl}$ is the solution of the longitudinal
equation of motion, and is a purely fluctuating field: that is,
it is fully determined by the initial charge fluctuations ${\cal W}$
(since ${\cal A}_L=0$ and ${\cal E}_L$ ix fixed by Gauss' law
(\ref{GAB1})). A calculation similar to the one above
eventually yields:
\beq\label{ZLA}
Z_{cl}[J_L]\,=\,
\exp\biggl\{\frac{T}{2}
\int\frac{{\rm d}^4k}{(2\pi)^4}\,\frac{\rho_L(k)}{k_0}\,\,
\frac{k^2}{k_0^2}\,\,|J_L(k)|^2
\biggr\}.\eeq
Note that the would-be pole at $k_0\to 0$ (as typical for the temporal
gauge) is actually innocuous in the classical theory, where
the current conservation $\del_\mu J^\mu=0$ ensures that the longitudinal
current also vanishes as $k_0\to 0$: $J_L(k_0,k)=(k_0/k) J_0(k_0,k)$.

Together, Eqs.~(\ref{ZTA}) and (\ref{ZLA}) provide the Abelian
partition function presented in Eqs.~(\ref{ZAB})--(\ref{DCL}).

\end{document}